\documentclass[a4paper,fleqn,usenatbib]{mnras}

\usepackage[T1]{fontenc}
\usepackage{ae,aecompl}

\usepackage{graphicx}	
\usepackage{amsmath}	
\usepackage{amssymb}	
\usepackage{multirow}
\usepackage{bm}

\newcommand{\dd}{\mathrm{d}}
\newcommand{\be}{\begin{equation}}
\newcommand{\ee}{\end{equation}}
\newcommand{\ba}{\begin{eqnarray}}
\newcommand{\ea}{\end{eqnarray}}

\title[Measuring the ISW effect with next-generation radio surveys]{Measuring the  ISW effect with next-generation radio surveys}

\author[Ballardini \& Maartens]{Mario Ballardini$^{1,2}$
\thanks{E-mail: \href{mailto:mario.ballardini@gmail.com}{mario.ballardini@gmail.com}}, 
Roy Maartens$^{1,3}$
\thanks{E-mail: \href{mailto:roy.maartens@gmail.com}{roy.maartens@gmail.com}}
\\
$^{1}$Department of Physics \& Astronomy, University of the Western Cape, Cape Town 7535, South Africa \\
$^{2}$INAF/OAS Bologna, via Gobetti 101, I-40129 Bologna, Italy \\
$^{3}$Institute of Cosmology \& Gravitation, University of Portsmouth, Portsmouth PO1 3FX, UK}

\pubyear{2019}

\begin{document}
\label{firstpage}
\pagerange{\pageref{firstpage}--\pageref{lastpage}}
\maketitle

\begin{abstract}
The late-time integrated Sachs-Wolfe (ISW) signal in the CMB temperature anisotropies is an 
important probe of dark energy when it can be detected by cross-correlation with large-scale 
structure surveys.  Because of their huge sky area, surveys in the radio are well-suited to 
ISW detection.
We show that 21cm intensity mapping and radio continuum surveys with the SKA in Phase~1 promise 
a $\sim5\sigma$ detection if we use tomography, with a similar forecast for the precursor EMU survey. 
In SKA Phase~2, the 21cm galaxy redshift survey and the continuum survey could deliver a 
$\sim6\sigma$ detection.
Our analysis of the radio surveys aims for theoretical accuracy on large scales. Firstly, we 
include all the effects on the radio surveys from observing on the past lightcone: 
redshift-space distortions and lensing magnification can have a significant impact on the ISW 
signal to noise ratio, while Doppler and other relativistic distortions are not significant. 
Secondly, we use the full information in the observable galaxy angular power spectra $C_\ell(z,z')$, 
by avoiding the Limber approximation and by including all cross-correlations between redshift bins 
in the covariance. Without these cross-bin correlations, the ISW signal to noise ratio is biased.
\end{abstract}




\section{Introduction}
\label{sec:intro}
The late-time integrated Sachs-Wolfe (ISW) effect  in the cosmic microwave background (CMB) 
arises from the time variation of the gravitational potentials along the line of sight 
\citep{Sachs:1967er}.
CMB photons are gravitational redshifted while travelling through potential wells and hills 
connected to matter over- and under-densities. In a matter-dominated universe the local 
gravitational potentials are constant and the net effect of a photon falling into a gravity 
well and coming out is zero. By contrast, gravitational potentials decay during the 
$\Lambda$-dominated phase, leading to a net change in photon temperature and in the observed 
CMB temperature anisotropies.
These potential fluctuations are induced by density perturbations at relatively low redshifts 
and generate a non-vanishing cross-correlation component between CMB temperature anisotropies 
and CMB lensing, and between CMB temperature anisotropies and number count fluctuations.

Many measurements of the ISW signal based on cross-correlating the CMB with large-scale 
structure tracers have been performed \citep{Crittenden:1995ak,Nolta:2003uy,Fosalba:2003iy,Boughn:2003yz,Afshordi:2003xu,Vielva:2004zg,McEwen:2006my,Ho:2008bz,Giannantonio:2008zi,Sarkar:2008sz,HernandezMonteagudo:2009fb,Taburet:2010hb,Ilic:2011hh,Schiavon:2012fc,Barreiro:2012yh,Ade:2013dsi,Ferraro:2014msa,Manzotti:2014kta,Ade:2015dva,Shajib:2016bes,Bianchini:2016czu,Stolzner:2017ged,Maniyar:2018hfp}.
Measurements have used different matter tracers: radio source catalogues, spectroscopic and 
photometric galaxy surveys, photometric quasars, thermal Sunyaev-Zeldovich, cosmic infrared 
background, and CMB lensing.
Alternatively, the stacking of CMB data in correspondence with superclusters and supervoids 
has also been used for ISW detection \citep{Granett:2008ju,Papai:2010gd,Ilic:2011hh,Ade:2013dsi,Ade:2015dva}.

ISW fluctuations contribute mainly to large angular scales,  $\ell \lesssim 100$, of the CMB 
temperature angular power spectrum, since there is little power in the potentials at late times 
on scales that entered the Hubble radius during radiation domination \citep{Kofman:1985fp}.
For this reason, wide surveys are optimal to tackle ISW detection. 
Such surveys can probe the large-scale structure on ultra-large (super-equality) scales, 
facilitating measurements of not only the ISW, but also primordial non-Gaussianity, the 
primordial power spectrum and relativistic observational effects on number counts and 
intensity \citep{Raccanelli:2011pu,Maartens:2012rh,Raccanelli:2014kga,Camera:2014bwa,Raccanelli:2015vla,Alonso:2015uua,Fonseca:2015laa,Ballardini:2017qwq,Karagiannis:2018jdt,Ballardini:2018noo,Bernal:2018myq}.
Among the next-generation surveys of the large-scale structure, radio surveys promise to 
deliver the largest volumes, 
using neutral hydrogen (HI) 21cm emission or radio continuum emission of galaxies 
\citep{Maartens:2015mra}. Radio surveys can maximise the synergies with CMB maps, thanks 
to their large overlapping sky area. 


In this paper, we test the feasibility of detecting the ISW signal with future cosmological 
surveys in Phase 1 of the Square Kilometre Array (SKA) \citep{Bacon:2018dui}, together with 
two of its precursor surveys, MeerKLASS \citep{Santos:2017qgq} and EMU \citep{Norris:2011ai}. 
We also consider the more futuristic `billion galaxy' spectroscopic survey and continuum survey 
in SKA Phase~2.

We begin by quantifying the theoretical signal-to-noise ratio (SNR) for  ISW detection through 
angular cross-power spectra of CMB temperature and number count/ intensity. 
Then we simulate ISW and large-scale structure (LSS) maps to identify the quality of the 
reconstruction for the radio surveys considered. 
We explore the impact of lensing magnification and other relativistic effects in LSS maps 
on the cross-correlation with CMB temperature maps \citep[see also][]{LoVerde:2006cj,Challinor:2011bk,Renk:2016olm}. 
When these effects are not modelled, we use the incorrect theoretical model, potentially 
leading to a bias in the ISW reconstruction. In addition, we highlight the importance of the 
cross-correlations between redshift bins when a tomographic approach is used. Without these 
cross-bin correlations, the covariance is not correctly modelled, leading to a bias in the 
ISW signal to noise.

The paper is organized as follows. We review the angular cross-power spectrum in 
section~\ref{sec:two}, and the SNR calculation in section~\ref{sec:three}.
In section~\ref{sec:four}, we describe the properties of the surveys.
Section~\ref{sec:five} discusses our results for the SNR analysis.
Our procedure for ISW reconstruction is described in section~\ref{sec:six}, together with 
discussion on the accuracy of using the average correlation coefficient and on map residuals 
between the true ISW map and the reconstructed one in pixel space. 
We draw our conclusions in section~\ref{sec:conclusion}.

\section{CMB-LSS cross-correlation}
\label{sec:two}

We study the cross-correlation between the LSS angular power spectrum derived in linear 
perturbation theory assuming general relativity (GR) and the ISW contribution to the CMB 
temperature angular power spectrum. There is a simple way to relate number count results 
to intensity mapping, which we describe below. Therefore in this section we consider the 
observed number counts for a magnitude-limited survey. 

In Newtonian gauge and in Fourier space, the multipoles of the observed number counts can 
be split into a standard term, which includes RSD by convention, and the GR contributions 
which are effectively the corrections to the standard Newtonian approximation:
\be
\label{eqn:counts}
\Delta_{\ell} ({\bm{k}},z) = \Delta^{\rm N}_\ell(\bm{k},z) + \Delta^{\rm GR}_\ell(\bm{k},z) \,.
\ee
The Newtonian term is the number density contrast plus the standard RSD term:
\begin{align}
\label{eqn:countsNewt}
\Delta^{\rm N}_\ell 
&= b\,{\delta^{\rm c}_{\bm{k}}}\, j_\ell(k\chi) 
+ \frac{k \,v_{\bm{k}}}{{\cal H}}\, j_\ell''(k\chi) \,.
\end{align}
Here 
$\delta^{\rm c}_{\bm{k}}(z)$ is the comoving matter density contrast, used in order to impose 
a physical model of scale-independent bias $b(z)$ \citep{Challinor:2011bk,Bruni:2011ta,Jeong:2011as,Baldauf:2012hs}. 
The peculiar velocity of the source is $\bm{v}$, with $v_{\bm{k}}(z)=|\bm{v}_{\bm{k}}(z)|$, 
and $j_\ell$ are spherical Bessel functions, where prime denotes $\dd/\dd (k\chi)$.
${\cal H} (z)= (1+z)^{-1}H(z)$ is the conformal Hubble parameter and $\chi(z)$ is the comoving 
distance. 

The GR corrections to $\Delta^{\rm N}$ are given by \citep{Challinor:2011bk}:\\
$\bullet$ the lensing convergence contribution (L), which is 
$\propto \delta^{\rm c}_{\bm{k}}$; \\
$\bullet$ the Doppler effect due to redshift perturbations from peculiar velocity (V), which 
is $\propto ({\cal H}/k)\delta^{\rm c}_{\bm{k}}$; \\
$\bullet$ ultra-large scale terms (ULS), which are 
$\propto ({\cal H}/k)^2 \delta^{\rm c}_{\bm{k}}$, and come from the  gravitational potentials:
\be
\dd s^2 = \left[-\left(1+2\psi\right)\dd \eta^2 + \left(1-2\phi\right)\dd \bm{x}^2\right] \,.
\ee

In detail \citep{Fonseca:2015laa}:
\begin{align}
\label{eqn:countsGR}
\Delta^{\rm GR}_\ell 
&\equiv \Delta^{\rm L}_\ell + \Delta^{\rm V}_\ell + \Delta^{\rm ULS}_\ell \,,\\ \label{eqn:countsL}
\Delta^{\rm L}_\ell 
&={ \ell(\ell+1) \over 2}(2-5s)  \nonumber\\  
&~\times \int_0^{\chi} \dd\tilde\chi \frac{\big(\chi-\tilde\chi\big)}{\chi\tilde\chi} 
\left[ \psi_{\bm{k}}(\tilde\chi)+\phi_{\bm{k}}(\tilde\chi)\right] j_\ell(k\tilde\chi)\,, \\
\label{eqn:countsV}
\Delta^{\rm V}_\ell 
&=\left[ \frac{2-5s}{{\cal H}\chi} + 5s - b_e + \frac{\dot{\cal H}}{{\cal H}^2} \right] v_{\bm{k}}\, j_\ell'(k\chi) \,,
\end{align}
where 
\begin{align}
s(z,m_*) &= {\partial \log \bar{\cal N}(z,m\!<\!m_*) \over \partial m_*}\,,\\ 
b_e(z,m_*) &=- {\partial \ln[(1+z)^{-3} \bar{\cal N}(z,m\!<\!m_*)] \over \partial \ln (1+z)}\,,
\end{align}
are the magnification and evolution bias, and $\bar{\cal N}$ is the background number density 
of sources with luminosity above the threshold.

The ULS contribution is made up of local and integrated terms:
\begin{align}
\label{eqn:countsULS} 
{\Delta^{\rm ULS}_\ell} 
&= \left\{\left[ \frac{2-5s}{{\cal H}\chi} + 5s - b_e + \frac{\dot{\cal H}}{{\cal H}^2} + 1\right]
\psi_{\bm{k}} \right. + \notag\\
&\left.  (5s-2)\phi_{\bm{k}} + \frac{\dot{\phi}_{\bm{k}}}{{\cal H}}+ {\left(b_e-3\right){\cal H}\frac{v_{\bm{k}}}{k}} \right\} j_\ell(k\chi) + \notag \\
&\left[ \frac{2-5s}{{\cal H}\chi} + 5s - b_e + \frac{\dot{\cal H}}{{\cal H}^2} \right] 
\!\int_0^{\chi}\! \dd\tilde\chi \left[\dot\psi_{\bm{k}}(\tilde\chi)+\dot\phi_{\bm{k}}(\tilde\chi)\right] j_\ell(k\tilde\chi) 
\notag\\
&+\frac{(2-5s)}{\chi} 
\int_0^{\chi} \dd\tilde\chi \left[\psi_{\bm{k}}(\tilde\chi)+\phi_{\bm{k}}(\tilde\chi)\right] j_\ell(k\tilde\chi) \,.
\end{align}
The first two lines of  \eqref{eqn:countsULS} are local Sachs-Wolfe type terms.
The velocity term $v_{\bm{k}}/{k}$ -- which  arises from expressing the Newtonian-gauge number 
density contrast in comoving gauge -- can be rewritten as a potential term using the Poisson 
and continuity equations \citep{Fonseca:2018hsu}. The last two lines are nonlocal integrated terms, 
from the ISW   and time-delay effects in the LSS density contrast.

The angular power spectra of the LSS density contrast and the CMB ISW are (suppressing the 
redshift dependence):
\be
\label{eqn:APS}
C_{\ell}^{\rm XY} = 4 \pi \int \frac{\dd k}{k} {\cal P_R} (k) I_\ell^{\rm X} (k) I_\ell^{\rm Y} (k) \,, 
\ee
where  X,Y = $\Delta$ or ISW and we use the convention $C_{\ell}^{\rm XX}=C_{\ell}^{\rm X}$. 
In \eqref{eqn:APS}, ${\cal P_R}$ is the dimensionless primordial power spectrum and the kernels are: 
\begin{align}
{I_\ell^{\Delta} (k,z_{\rm i})} &= \int \dd z \, {W(z,z_{\rm i})}\, {\Delta_\ell} (k,z) \,,\\
I_\ell^{\rm ISW} (k) &= \int \dd z \, e^{-\tau(z)} 
{\mathcal{T_{ \dot{\phi}+\dot{\psi}}}(k,z)}\, j_\ell (k \chi(z)) \,.
\end{align}
Here $W(z,z_{\rm i})$ is the window function for the LSS redshift bin $z_{\rm i} \pm \Delta z/2$,
$\tau$ is the optical depth, and $\Delta_\ell (k,z)$ is obtained from \eqref{eqn:countsNewt}, \eqref{eqn:countsL}, \eqref{eqn:countsV}, \eqref{eqn:countsULS} through replacing 
$\delta_{\bm{k}}^{\rm c}(z)$  by its transfer function $\mathcal{T}_{\delta^{\rm c}}(k,z)$, 
and similarly for $v_{\bm{k}},\psi_{\bm{k}},\phi_{\bm{k}}$.
We use top-hat windows for the HI surveys (which have excellent spectroscopic accuracy), and  
Gaussian windows for the continuum survey.
We do not use the Limber approximation, which is not accurate on the large scales where the 
ISW signal is strongest.

\section{Theoretical signal-to-noise} 
\label{sec:three}
The observed LSS auto-power spectra are
\be
\bar{C}_\ell^{\Delta}(z_{\rm i})={C}_\ell^{\Delta}(z_{\rm i}) +
{\cal N}_\ell(z_{\rm i}) \,,
\ee
where ${\cal N}_\ell$ is shot noise for galaxies and thermal noise for
intensity mapping (see below).

In order to quantify the possibility of extracting the ISW signal from the CMB, we follow 
\citep{Cooray:2001ab,Afshordi:2004kz} and use the signal-to-noise ratio (SNR) defined by:
\be
\label{eqn:SNR}
\left(\frac{S}{N}\right)^2 =
\sum_{\ell=2}^{\ell_{\rm max}}\left({\bf C}_\ell^{\Delta\, {\rm ISW}}\right)^\dagger
\Big({\bf C}_\ell^{\rm cov}\Big)^{-1} 
{\bf C}_\ell^{\Delta\, {\rm ISW}} \,.
\ee
Here ${\bf C}_\ell^{\Delta\, {\rm ISW}}(z_{\rm i})$ is the vector of the angular 
cross-power spectra, and the covariance matrix elements are:
\be \label{covm}
{\bf C}_\ell^{\rm cov}(z_{\rm i},z_{\rm j}) =
\frac{C_\ell^{\Delta\, {\rm ISW}}(z_{\rm i})\,C_\ell^{\Delta\, {\rm ISW}}(z_{\rm j}) + \bar{C}_\ell^{\Delta}(z_{\rm i},z_{\rm j})\,\bar{C}_\ell^{\rm ISW}}{\left(2\ell+1\right)f_{\rm sky}} , 
\ee
where $f_\textrm{sky}$ is the common sky fraction of the LSS and CMB surveys. 
Cross-bin LSS correlations $\bar{C}_\ell^{\Delta}(z_{\rm i},z_{\rm j})$, i$\,\neq\,$j, do not 
enter the ISW signal ${\bf C}_\ell^{\Delta\, {\rm ISW}}$, but they do affect the SNR, via the 
covariance matrix \eqref{covm}. Neglecting these cross-bin correlations can therefore bias the SNR.

For the ISW contribution to the CMB, the primary temperature auto-power is part of the noise:
\be
\label{eqn:ISWnoise}
\bar{C}_\ell^{\rm ISW} = {C}_\ell^{\rm ISW} + C_\ell^\mathrm{TT} + {\cal N}_\ell^\mathrm{T} \,.
\ee
At low and intermediate multipoles, where the ISW signal is not suppressed, the instrumental 
noise on the CMB temperature signal can be neglected, i.e. ${\cal N}_\ell^{\rm T}\approx 0$.

The CMB E-mode polarization can be used to improve the ISW signal in the CMB temperature 
angular power spectrum, thanks to the TE correlation between the two spectra.
By including E-mode polarization information, it is possible to decrease the effective 
cosmic variance on the CMB temperature and ISW angular power spectra:
\be
\label{eqn:ISWnoiseTE}
\bar{C}_\ell^{\rm ISW} = C_\ell^{\rm ISW} + C_\ell^\mathrm{TT} 
- \frac{\left(C_\ell^\mathrm{TE}\right)^2}{C_\ell^\mathrm{EE}} + {\cal N}_\ell^\mathrm{T} \,.
\ee
The inclusion of  CMB polarization in the analysis increases the SNR by  $\sim\!18\%$ 
\citep{Frommert:2008qh,Giannantonio:2012aa,Ballardini:2017xnt}. Therefore in the SNR 
calculation we always  include the E-mode polarization information according to 
\eqref{eqn:ISWnoiseTE}.

\section{Radio survey specifications}
\label{sec:four}
In this section we provide the LSS survey details used for the analysis. 

\subsection{HI intensity mapping survey}
\label{sec:im}
HI intensity mapping (IM) surveys do not attempt to detect individual HI galaxies, but measure 
the total signal in each pixel  to produce maps of the large-scale fluctuations in HI galaxy 
clustering (with very accurate redshifts) \citep{Battye:2004re,Wyithe:2007gz,Chang:2007xk,Bull:2014rha,Santos:2015bsa,Kovetz:2017agg}.
The flux density measured is converted into an effective brightness temperature of the 21cm emission:
\be \label{imt}
T_{{\rm HI}}=\bar{T}_{{\rm HI}}\big(1+\delta_{\rm HI}\big)\ \mu\text{K} \,.
\ee

HI is expected to be a biased tracer of the cold dark matter distribution, just as galaxies 
are, because the HI content of the Universe is expected to be localized within galaxies 
after reionization. We use the fitting formulas \citep{Santos:2017qgq}:
\begin{align}
b_{\rm HI}(z) &= \frac{b_{{\rm HI}}(0)}{0.677105} \Big[0.66655 + 0.17765\, z + 0.050223\,  z^2\Big]\!,  \\
\bar{T}_{{\rm HI}}(z) &= 0.055919 + 0.23242\, z - 0.024136\, z^2\ \text{mK} \,,
\end{align}
where $\Omega_{\rm HI}(0)b_{\rm HI}(0) = 4.3\times 10^{-4}$ and 
$\Omega_{\rm HI}(0) = 4.86\times 10^{-4}$.

The observed brightness temperature contrast may be obtained from the number count case by using 
effective values for the evolution and magnification biases as follows \citep{Hall:2012wd,Fonseca:2018hsu}:
\be \label{hibes}
b_{e{\rm HI}}(z)=-{\partial \ln \big[(1+z)^{-1}{\cal H}(z)\bar{T}_{\rm HI}(z) \big] \over \partial \ln(1+z)}\,,\quad s_{{\rm HI}}= {2\over 5}\,.
\ee
Note that the lensing magnification contribution is thus zero at first order.

We consider IM in single-dish mode, i.e. adding up all dishes independently as opposed to 
combining them via interferometry, using SKA1-MID Band 1 and the proposed  IM survey MeerKLASS 
on the precursor MeerKAT. Single-dish mode is the most efficient way to probe cosmological scales 
with IM \citep{Santos:2015bsa}. 
Assuming scale-independence and no correlation between the noise in different frequency channels, 
the noise variance per steradian in the $i$-frequency channel is \citep{Knox:1995dq,Bull:2014rha}:
\ba
\label{eqn:Nhi}
{{\cal N}_\ell^{\rm HI}}(\nu_{\rm i}) &=& \frac{4\pi f_{\rm sky}T^2_{\rm sys}(\nu_{\rm i})}{2N_{\rm dish}t_{\rm tot}\Delta\nu} \,,\\
T_{\rm sys}(\nu) &=& 25 + 60\left(\frac{300\,\text{MHz}}{\nu}\right)^{2.55}\ \text{K} \,.
\ea

For MeerKLASS, we assume $N_{\rm dish} = 64$, $t_{\rm tot} = 4,000$ hr over 4,000\,deg$^2$ 
($f_{\rm sky} \simeq 0.1$) in the redshift ranges $0\le z\le 0.58$ ($1670\geq\nu\geq900\,$MHz, 
L Band) and $0.4\le z\le 1.45$ ($1015\geq\nu\geq580\,$MHz, UHF Band) \citep{Santos:2017qgq}. 

For SKA1-MID, we assume $N_{\rm dish} = 197$, $t_{\rm tot} = 10^4$ hr observing over 
20,000\,deg$^2$ ($f_{\rm sky} \simeq 0.5$) in the redshift range $0.35 \le z \le 3.05$ 
($1050 \geq \nu \geq 350\,$MHz, Band 1) \citep{Bacon:2018dui}.

\subsection{Radio continuum survey}
\label{sec:cont}
A continuum survey detects the total radio emission of galaxies, which is dominated by 
synchrotron radiation. As a consequence, there is no redshift information, and redshifts must 
be obtained by cross-matching with optical/ infrared surveys (or HI IM surveys). On the other 
hand, radio galaxies are detected out to very high redshift.

For a continuum galaxy survey with SKA1-MID, we assume the same  frequency band and sky area 
as for the IM survey, with a source detection limit $S_{\rm cut} = 22\,\mu$Jy 
\citep{Bacon:2018dui}. 
We study also an optimistic case for SKA1-MID with $S_{\rm cut} = 10\,\mu$Jy.
For SKA2 we assume 30,000\,deg$^2$ ($f_{\rm sky} \simeq 0.7$) and $S_{\rm cut} = 1\,\mu$Jy.

As an SKA precursor continuum galaxy survey, we consider EMU on ASKAP 
\citep{Norris:2011ai,Bernal:2018myq}, covering the frequency range 700--1450\,MHz, 
with $f_{\rm sky} \simeq 0.7$ and $S_{\rm cut} = 100\,\mu$Jy.

The redshift distribution, bias, magnification bias and evolution bias are predicted using 
the publicly available code\footnote{\href{http://intensitymapping.physics.ox.ac.uk/codes.html}{http://intensitymapping.physics.ox.ac.uk/codes.html}}
developed by \cite{Alonso:2015uua}, which provides semi-analytical fits based on the 
simulations described in \cite{Wilman:2008ew}.

\subsection{HI galaxy survey}
\label{sec:gal}

The models for the number counts and the clustering, evolution and magnification biases 
of the HI galaxy distribution are given in \cite{Camera:2014bwa}, for various flux thresholds.

For SKA1, the sky area and redshift coverage are too low for detecting the ISW. SKA2 
specifications have not been formalised, but the rms noise is expected to be $\sim\!10$ 
times smaller than SKA1, and the sky coverage increased from 20,000 to 30,000\,deg$^2$ 
($f_{\rm sky} \simeq 0.7$). We assume a total observation time of $10^4\,$hr 
 in a redshift range $0.1\le z\le 2$, and with flux threshold $1\,\mu$Jy. 
This is the so-called `billion galaxy' survey.\\

Note that plots of the number density, clustering bias, evolution bias, and magnification 
bias as functions of $z$ for the three types of radio survey can be found in \cite{Camera:2014bwa} 
and \cite{Alonso:2015uua}.

\section{Results}
\label{sec:five}

\subsection{HI IM survey} 

\begin{table*}
\centering
\caption{SNR for ISW detection from IM surveys with MeerKAT and SKA1-MID, using binning 
configurations \eqref{zb1} and \eqref{zb2}. SNR is calculated using the full HI temperature 
contrast $\Delta^{\rm N}+\Delta^{\rm GR}$ given by \eqref{eqn:counts} (left column) and its 
Newtonian approximation $\Delta^{\rm N}$ given by \eqref{eqn:countsNewt} (right column). 
Numbers in round brackets indicate {\em neglect} of the contribution from cross-bin correlations in \eqref{eqn:SNR}.}
\label{tab:SNim}
\begin{tabular}{|c|cc|cc|cc|}
\hline
\rule[-1.2mm]{0mm}{.45cm}
 & \multicolumn{2}{|c|}{MeerKLASS}  & \multicolumn{2}{|c|}{MeerKLASS} & \multicolumn{2}{|c|}{SKA1-MID} \\
 & \multicolumn{2}{|c|}{L-band} & \multicolumn{2}{|c|}{UHF-band} & \multicolumn{2}{|c|}{Band 1} \\
\hline
\rule[-1.2mm]{0mm}{.45cm}
& $\Delta$-ISW & $\Delta^{\rm N}$-ISW & $\Delta$-ISW & $\Delta^{\rm N}$-ISW & $\Delta$-ISW & $\Delta^{\rm N}$-ISW \\
\hline
\rule[-1.2mm]{0mm}{.45cm}
1 bin & $1.1$ & $1.1$ & $1.7$ & $1.7$ & $3.7$ & $3.7$ \\
\rule[-1.2mm]{0mm}{.45cm}
$\Delta z=0.1$ bins & $1.3\ (1.2)$ & $1.3\ (1.2)$ & $1.9\ (1.6)$ & $1.9\ (1.6)$ & $4.7\ (3.9)$ & $4.7\ (3.9)$ \\
\hline
\end{tabular}
\end{table*}

Table~\ref{tab:SNim} summarises the SNR obtained by correlating the observed HI 
brightness temperature contrast $\Delta_{\rm HI}$ with the CMB ISW signal. We calculate the 
SNR in two different configurations, for MeerKAT L-band/ UHF-band/ SKA1-MID Band 1 respectively:
\begin{align}
\quad &1~ z\mbox{-bin with edges } [0, 0.58]/ [0.4, 1.45]/ [0.35, 3.05] \label{zb1} \\ 
~& 5/11/27~ z\mbox{-bins with } \Delta z=0.1  \label{zb2}
\end{align}
In each case, we compare the SNR using the Newtonian approximation $\Delta^{\rm N}$ with the 
SNR using the full GR result $\Delta=\Delta^{\rm N}+\Delta^{\rm GR}$. For the cases with 
tomography, we also show the effect (in brackets) of {\em neglecting} cross-bin correlations 
in the brightness temperature.

The SNR for MeerKLASS IM surveys is strongly limited by the survey area of 4,000\,deg$^2$. 
Moreover, the instrumental noise limits the possibility to increase the SNR by considering 
many slices in redshift. We find a SNR $\sim 1.1-1.3$ for the L-band and $\sim 1.7-1.9$ for 
the UHF-band. 
For an SKA1-MID Band 1 IM survey, we find a SNR of $\sim 3.7$ using the whole survey as one 
single redshift bin, increasing up to $\sim 4.7$ using tomography. 

\begin{figure*}
\centering
\includegraphics[width=14cm]{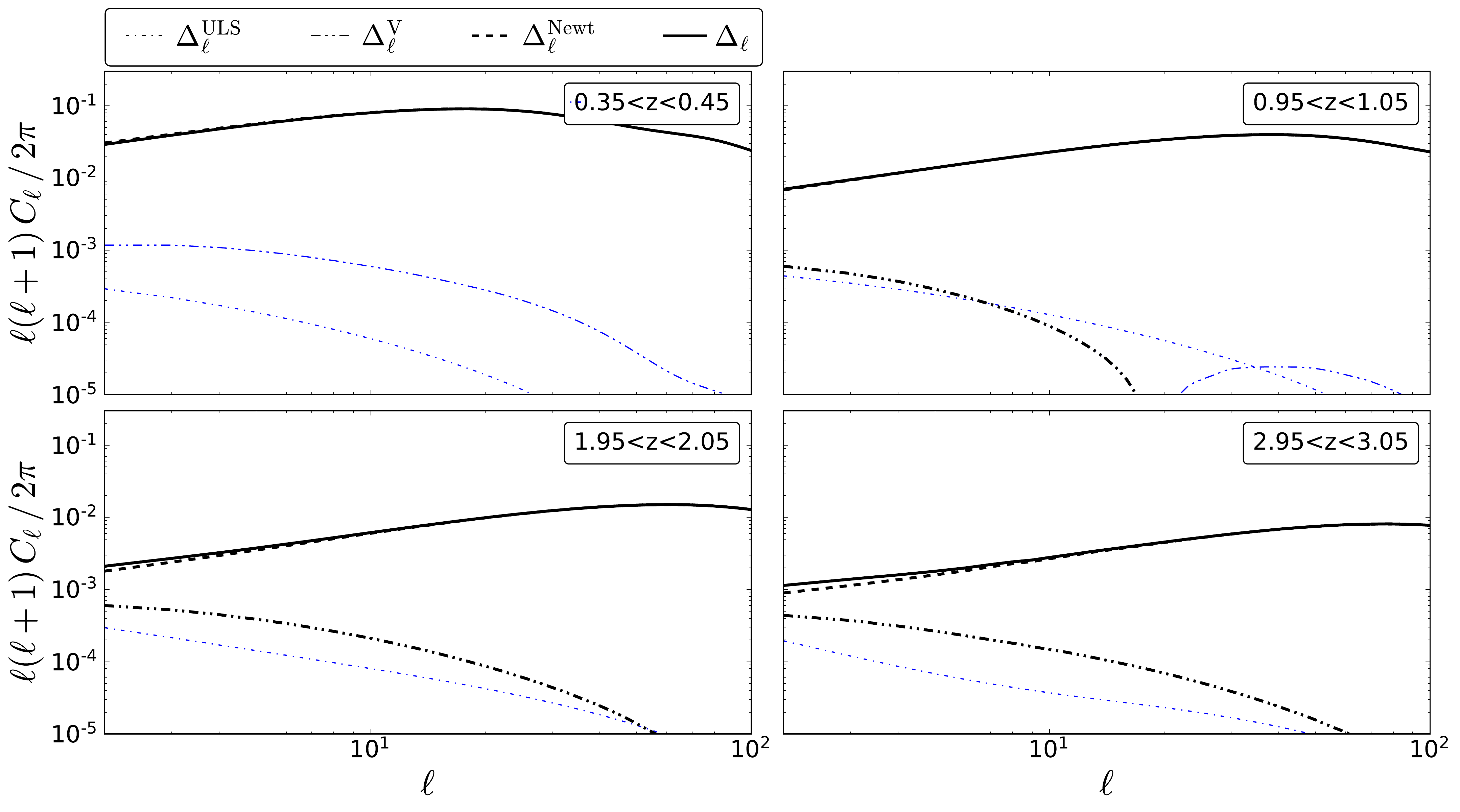}
\caption{HI temperature contrast-ISW cross-correlation for an SKA1-MID IM survey, showing 4 of 
the 27 redshift bins. Different lines correspond to the contributions from ultra-large scale GR 
effects ($\Delta^{\rm ULS}$, dot-dashed), from Doppler effects ($\Delta^{\rm V}$, dot-dot-dashed), 
from the Newtonian approximation ($\Delta^{\rm N}$, dashed), and from the total ($\Delta$, solid). 
Thin (blue) lines indicate the absolute value of a negative contribution.}
\label{fig:Tg_im4}
\end{figure*}

In IM, the lensing magnification contribution $\Delta^{\rm L}$ given by \eqref{eqn:countsL} 
is zero, as follows from \eqref{hibes}. 
As a further consequence of $s_{\rm HI} = 2/5$, parts the other GR terms, $\Delta^{\rm V}$ 
(Doppler) given by \eqref{eqn:countsV} and $\Delta^{\rm ULS}$ (potentials) given by 
\eqref{eqn:countsULS}, are suppressed.
The contribution $\Delta^{\rm GR}$ from the total GR correction in the observed temperature 
contrast $\Delta$ is much smaller than the Newtonian $\Delta^{\rm N}$ contribution, even at 
high redshift. This can be seen in figure~\ref{fig:Tg_im4}. 
Table~\ref{tab:SNim} shows that for IM there is a negligible difference between the SNR in 
full GR  and the  SNR in the Newtonian approximation. 

In all cases,  {\em tomography improves the SNR}. If we neglect the IM cross-bin correlations 
in the covariance \eqref{covm}, then the SNR is biased downward (numbers in brackets) -- i.e. 
cross-bin correlations increase the SNR. {\em Neglecting cross-bin correlations in IM leads to 
a false reduction in SNR}, due to incorrect modelling of the covariance.

In \cite{Pourtsidou:2016dzn}, the SNR for ISW detection is calculated for MeerKLASS UHF-band 
and SKA1-MID Band 1, assuming independent redshift bins with a width of $\Delta z = 0.1$. 
They therefore neglect cross-bin correlations, and they find a SNR of 1.5 and 4.6.
The difference with our results is likely due to their assumptions of a different best-fit 
cosmology, a simplified redshift independent bias, $b_{\rm HI}(z) = 1$, and (for SKA1) a 
larger sky coverage of 30,000\,deg$^2$.

\subsection{Radio continuum survey} 
\begin{table*}
\centering
\caption{SNR for ISW detection from continuum galaxy surveys with ASKAP and SKA1-MID, using 
binning configurations \eqref{zb3} and \eqref{zb4}. SNR is calculated using the full number 
count contrast $\Delta^{\rm N}+\Delta^{\rm GR}$ given by \eqref{eqn:counts} (left column) and 
its Newtonian approximation $\Delta^{\rm N}$ given by \eqref{eqn:countsNewt} (right column). 
Numbers in round brackets indicate {\em neglect} of the contribution from cross-bin correlations 
in \eqref{eqn:SNR}.}
\label{tab:SNcont}
\begin{tabular}{|c|cc|cc|cc|cc}
\hline
\rule[-1.2mm]{0mm}{.45cm}
 & \multicolumn{2}{|c|}{EMU}  & \multicolumn{2}{|c|}{SKA1-MID Band 1} & \multicolumn{2}{|c|}{SKA1-MID Band 1} & \multicolumn{2}{|c|}{SKA2} \\
 & \multicolumn{2}{|c|}{$S_{\rm cut} = 100$ $\mu$Jy} & \multicolumn{2}{|c|}{$S_{\rm cut} = 22$ $\mu$Jy} & \multicolumn{2}{|c|}{$S_{\rm cut} = 10$ $\mu$Jy} & \multicolumn{2}{|c|}{$S_{\rm cut} = 1$ $\mu$Jy} \\
\hline
\rule[-1.2mm]{0mm}{.45cm}
& $\Delta$-ISW & $\Delta^{\rm N}$-ISW & $\Delta$-ISW & $\Delta^{\rm N}$-ISW & $\Delta$-ISW & $\Delta^{\rm N}$-ISW & $\Delta$-ISW & $\Delta^{\rm N}$-ISW \\
\hline
\rule[-1.2mm]{0mm}{.45cm}
1 bin & $5.0$ & $5.0$ & $4.0$ & $4.0$ & $5.1$ & $5.1$ & $5.6$ & $5.6$ \\
\rule[-1.2mm]{0mm}{.45cm}
5 bins & $-$ & $-$ & $5.0\ (5.1)$ & $5.0\ (5.2)$ & $5.1\ (5.3)$ & $5.1\ (5.4)$ & $6.2\ (6.0)$ & $6.2\ (6.6)$ \\
\hline
\end{tabular}
\end{table*}

\begin{table*}
\centering
\caption{SNR for ISW detection for an SKA2 HI galaxy survey, using binning configurations 
\eqref{zb5} and \eqref{zb6}. SNR is calculated using the full number  count contrast 
$\Delta^{\rm N}+\Delta^{\rm GR}$ given by \eqref{eqn:counts} (left column) and its Newtonian 
approximation $\Delta^{\rm N}$ given by \eqref{eqn:countsNewt} (right column). 
Numbers in round brackets indicate {\em neglect} of the contribution from cross-bin correlations 
in \eqref{eqn:SNR}.}
\label{tab:SNgal}
\begin{tabular}{|c|cc|}
\hline
\rule[-1.2mm]{0mm}{.45cm}
 & \multicolumn{2}{|c|}{SKA2} \\
\hline
\rule[-1.2mm]{0mm}{.45cm}
& $\Delta$-ISW & $\Delta^{\rm N}$-ISW \\
\hline
\rule[-1.2mm]{0mm}{.45cm}
1 bin & $3.8$ & $3.7$ \\
\rule[-1.2mm]{0mm}{.45cm}
$\Delta z=0.1$ bins & $6.0\ (6.1)$ & $6.0\ (5.3)$ \\
\hline
\end{tabular}
\end{table*}
\begin{figure*}
\centering
\includegraphics[width=14cm]{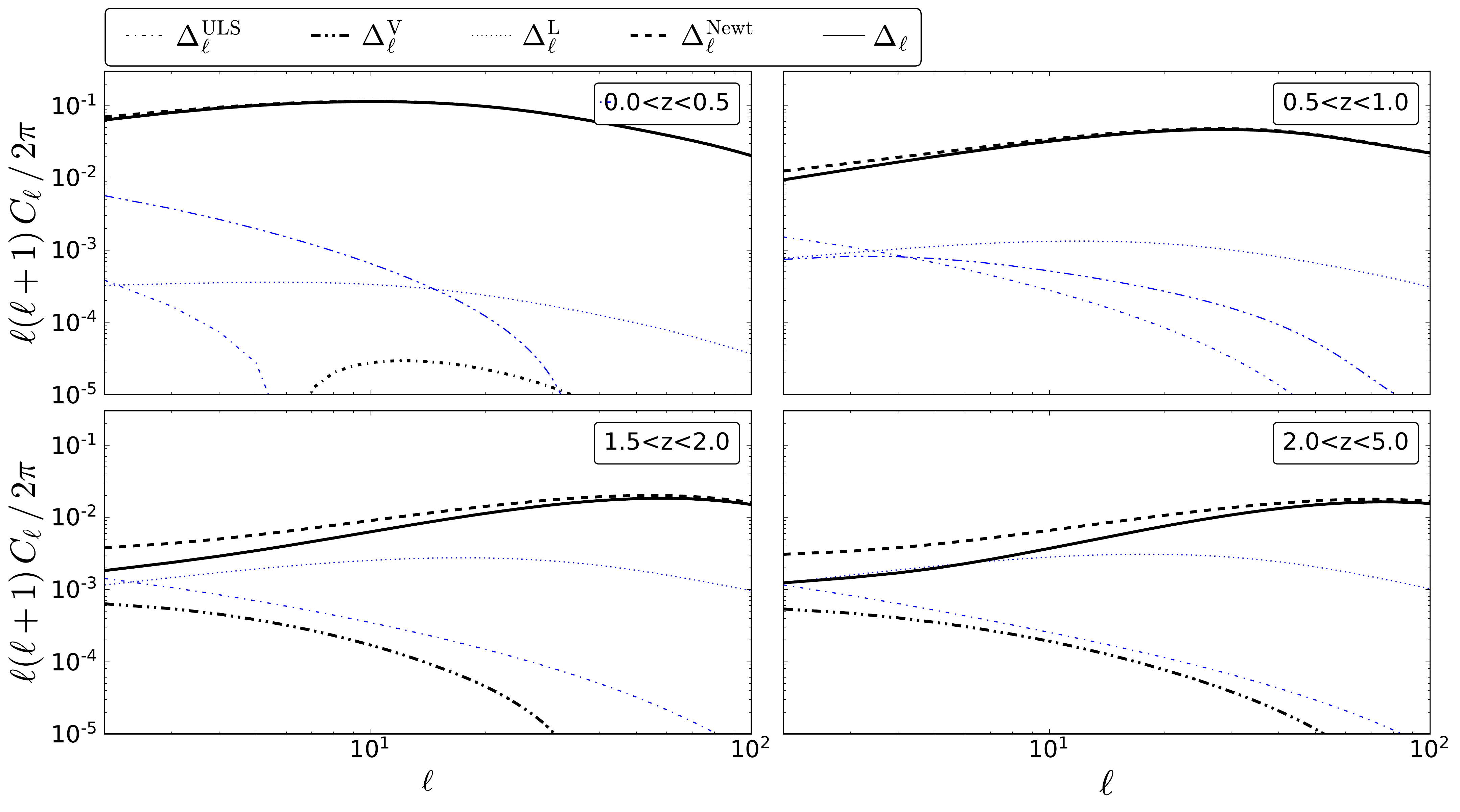}
\caption{As in figure~\ref{fig:Tg_im4}, for an SKA1 continuum survey ($S_{\rm cut} = 22\,\mu$Jy), 
showing 4 of the 5 redshift bins. The lensing contribution  (dotted), absent in 
figure~\ref{fig:Tg_im4}, is from $\Delta^{\rm L}$.}
\label{fig:Tg_cont4}
\end{figure*}
\begin{figure*}
\centering
\includegraphics[width=14cm]{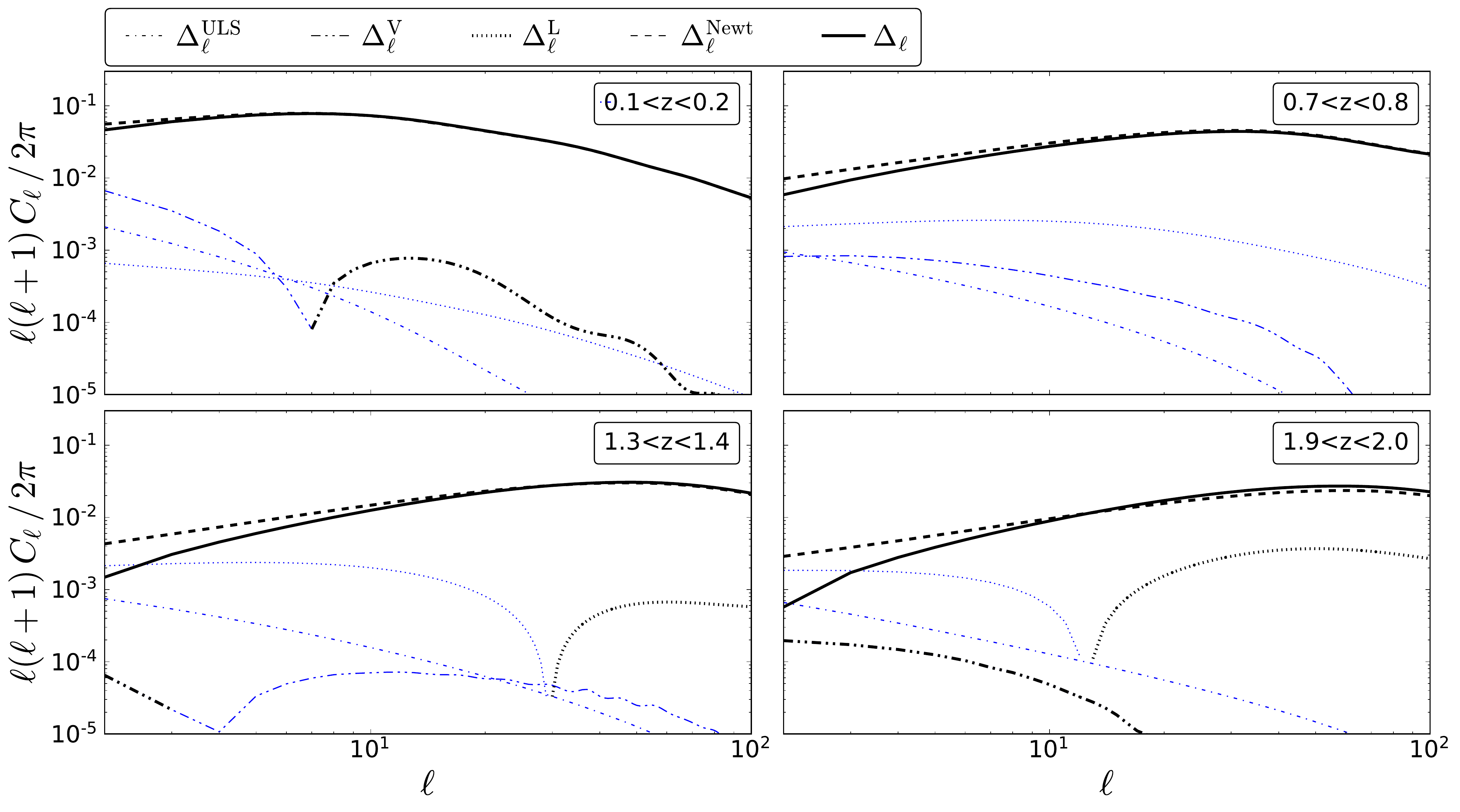}
\caption{As in figure~\ref{fig:Tg_cont4}, for an SKA2 HI galaxy survey, showing 4 of the 19 
redshift bins.}
\label{fig:Tg_gal4}
\end{figure*}
\begin{figure*}
\centering
\includegraphics[width=18cm]{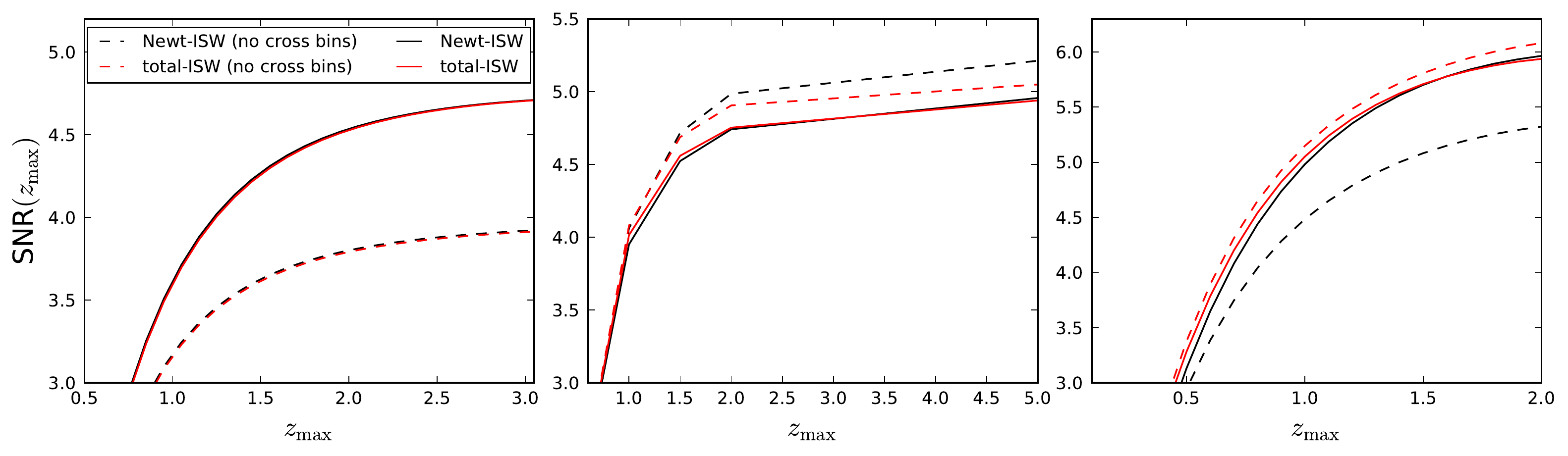}
\caption{Cumulative SNR  up to $z_{\rm max}$ for SKA1 IM (left panel), SKA1 continuum 
($S_{\rm cut}=22\,\mu$Jy) (central panel), and SKA2 HI galaxy survey (right panel). 
Different lines correspond to the cross-correlation $\Delta^{\rm N}$-ISW (red) and 
$\Delta$-ISW (black), with (solid) and without (dashed) cross-bin correlations.}
\label{fig:SNR}
\end{figure*}

In table~\ref{tab:SNcont}, we show the SNR obtained by correlating the observed number 
count contrast  with the CMB ISW, for EMU and SKA1 continuum surveys. 
We use the binning configurations:
\begin{align}
\quad &1~ z\mbox{-bin~ with edges } [0, 5] \label{zb3} \\ 
~& 5~ z\mbox{-bins with edges } [0, 0.5, 1, 1.5, 2, 5]  \label{zb4}
\end{align}
The second configuration applies only to SKA1 and is given in \cite{Bacon:2018dui}, based 
on the argument that sufficient spectroscopic information will be found (from cross-matching 
with 21cm and optical/ infrared surveys) to construct $\Delta z = 0.5$ bins up to $z = 2$.

Thanks to the wide sky coverage, we find a SNR of $\sim 5$ for EMU, despite its higher flux 
threshold. We obtain SNR $\sim 4.0-5.0$ for the SKA1 survey with the baseline 
$S_{\rm cut} = 22$ $\mu$Jy. For the optimistic case of $S_{\rm cut} = 10\,\mu$Jy, we find 
SNR $\sim 5.1$. For SKA2, with a larger sky area 30,000 deg$^2$ and lower flux threshold 
$1\,\mu$Jy, SNR $\sim 5.6-6.2$.
Similar to IM,  tomography improves the SNR, although the improvement is negligible in the 
$10\,\mu$Jy case.

Unlike IM surveys, continuum galaxy surveys are affected by lensing magnification. Figure~\ref{fig:Tg_cont4} 
shows that the lensing contribution  $\Delta^{\rm L}$-ISW (dotted line) to the full cross-power 
spectrum (solid line) is  important at high redshift and on the large scales where the ISW signal 
is stronger. It is of the same order of magnitude as the $\Delta^{\rm N}$-ISW contribution 
(dashed line) for the highest redshift bins. 

In figure~\ref{fig:Tg_cont4}, we also see that the correlation between $\Delta^{\rm L}$ 
and ISW is always negative (thin blue lines) and it is the dominant GR correction for 
$z \gtrsim 1$. 
This lensing contribution reduces the cross-power spectrum  relative to the Newtonian 
approximation $\Delta^{\rm N}$-ISW. 

Table~\ref{tab:SNcont} shows the following features in the tomographic case. If we neglect 
the galaxy cross-bin correlations in the covariance \eqref{covm}  (which gives a biased SNR), 
then the SNR is larger in the Newtonian approximation (no lensing) than in full GR (with lensing). 
This means that lensing effects {\em within} each bin reduce the SNR (consistent with 
figure~\ref{fig:Tg_cont4}).

The SNR without cross-bin correlations is also larger than the SNR in the cross-bin case, 
with the exception of the SKA2 case. This means that correlations amongst number count 
contrast, RSD and lensing  {\em across different} bins reduce the SNR, except for SKA2 in 
the full GR case, where there is an increase due to lensing.   {\em Neglecting cross-bin 
correlations in continuum leads to a false increase in SNR (a decrease for SKA2 in full 
GR) and a false excess SNR in the Newtonian approximation}, due to incorrect modelling 
of the covariance.

Finally, table~\ref{tab:SNcont} also shows that {\em when cross-bin correlations are included, 
the SNR in the Newtonian approximation (i.e. no lensing) differs negligibly from the SNR in the 
full GR model (including lensing).} This means that the total lensing contribution in 
\eqref{eqn:SNR}, from correlations of the form 
\begin{align}
\quad & \big\langle \Delta^{\rm L}(\bm{n},z_{\rm i})\, T^{\rm ISW}(\bm{n}')\big\rangle~\mbox{(within bins) and} \notag\\ 
&\big\langle \Delta^{\rm L}(\bm{n},z_{\rm i})\, \Delta(\bm{n}',z_{\rm j})\big\rangle~ \mbox{(within and across bins)}, \label{n=gr}
\end{align}
nearly cancels, so that neglecting lensing does not bias the SNR appreciably.

\subsection{HI galaxy survey}

Table~\ref{tab:SNgal} summarises the SNR obtained for detection of the correlation of  number 
count contrast and ISW  for an SKA2 HI galaxy survey. We use two binning configurations:
\begin{align}
\quad &1~~\, z\mbox{-bin~ with edges } [0.1, 2] \label{zb5} \\ 
~& 19~ z\mbox{-bins with  } \Delta z=0.1  \label{zb6}
\end{align}

Figure~\ref{fig:Tg_gal4} shows that the $\Delta^{\rm L}$-ISW contribution to the observed 
number count contrast-ISW cross-correlation is negative on the largest scales, similar to 
the continuum case, but can be positive at smaller scales (with $\ell<100$). 
Tomography gives a major boost to the SNR, which reaches $\sim6$ in the full model. 
Table~\ref{tab:SNgal} shows that if cross-bin correlations are neglected, the SNR is smaller 
in the Newtonian approximation -- which means that lensing effects {\em within} each bin increase 
the SNR. 
Without cross-bin correlations, the SNR in the full GR model is larger (6.1). This means that 
lensing effects {\em across} different bins reduce the SNR from its biased to true value (6.0). 
In other words, {\em neglecting cross-bin correlations in SKA2 galaxy surveys leads to a false 
increase in SNR and a false deficit of SNR in the Newtonian approximation}, due to incorrect 
modelling of the covariance.

Table~\ref{tab:SNgal} also shows that, as in the case of continuum, when cross-bin correlations 
are included, SNR in the Newtonian approximation (i.e. no lensing) differs negligibly from the SNR 
in the correct model (full GR, including lensing). It follows that correlations of the form 
\eqref{n=gr} again effectively cancel.



\subsection{Summary: cumulative SNR for SKA}

The cumulative SNR as a function of $z_{\rm max}$ for the baseline SKA1 surveys and the SKA2 HI 
galaxy survey is shown in figure~\ref{fig:SNR}. The plots illustrate the key features identified 
in the previous subsections. They also show that the SNR continues to grow at redshifts where 
the ISW is very small.



\section{Reconstructing the ISW signal}
\label{sec:six}
We use the optimal minimum-variance $\hat{a}^{\rm ISW}_{\rm \ell m}$ estimator derived 
in \cite{Barreiro:2008sn,Manzotti:2014kta} to reconstruct the ISW signal from CMB and LSS maps 
\citep[see also applications of the same estimators in][]{Muir:2016veb,Weaverdyck:2017ovf}:
\be
\label{eqn:estimator}
\hat{a}^{\rm ISW}_{\rm \ell m} = \sum_{i} R_\ell^i a_{\ell m}^i \,,
\ee
where $i =1,\cdots,n$ and the reconstruction filter derived from the covariance matrix is:
\be
R_\ell^i = -N_\ell \left(D_\ell^{-1}\right)_{1,i} \,.
\ee
The covariance matrix is:
\be
\label{eqn:cov}
D_\ell = 
\left( {\begin{array}{ccccc}
C_\ell^{\rm ISW} & C_\ell^{{\rm ISW} \Delta_1} & \dots & C_\ell^{\Delta_1{\rm ISW}} & C_\ell^{\rm ISW} \\
C_\ell^{\Delta_1{\rm ISW}} & \bar{C}_\ell^{\Delta_1} & \dots & \dots & C_\ell^{\Delta_1{\rm ISW}} \\
\vdots & \vdots & \ddots & \vdots & \vdots \\
C_\ell^{\Delta_n{\rm ISW}} & \dots & \dots & \bar{C}_\ell^{\Delta_n} & C_\ell^{\Delta_n{\rm ISW}} \\
C_\ell^{\rm ISW} & C_\ell^{{\rm ISW}\Delta_1} & \dots & \bar{C}_\ell^{\Delta_n} & C_\ell^{TT} \\
\end{array} } \right) 
\ee
Here the estimated variance of the reconstruction is:
\be
N_\ell = \big[(D_\ell^{-1})_{1,1}\big]^{-1} \,.
\ee
We reconstruct the ISW using the CMB temperature and $n$ LSS maps.

\begin{figure}
\centering
\includegraphics[width=.4\textwidth]{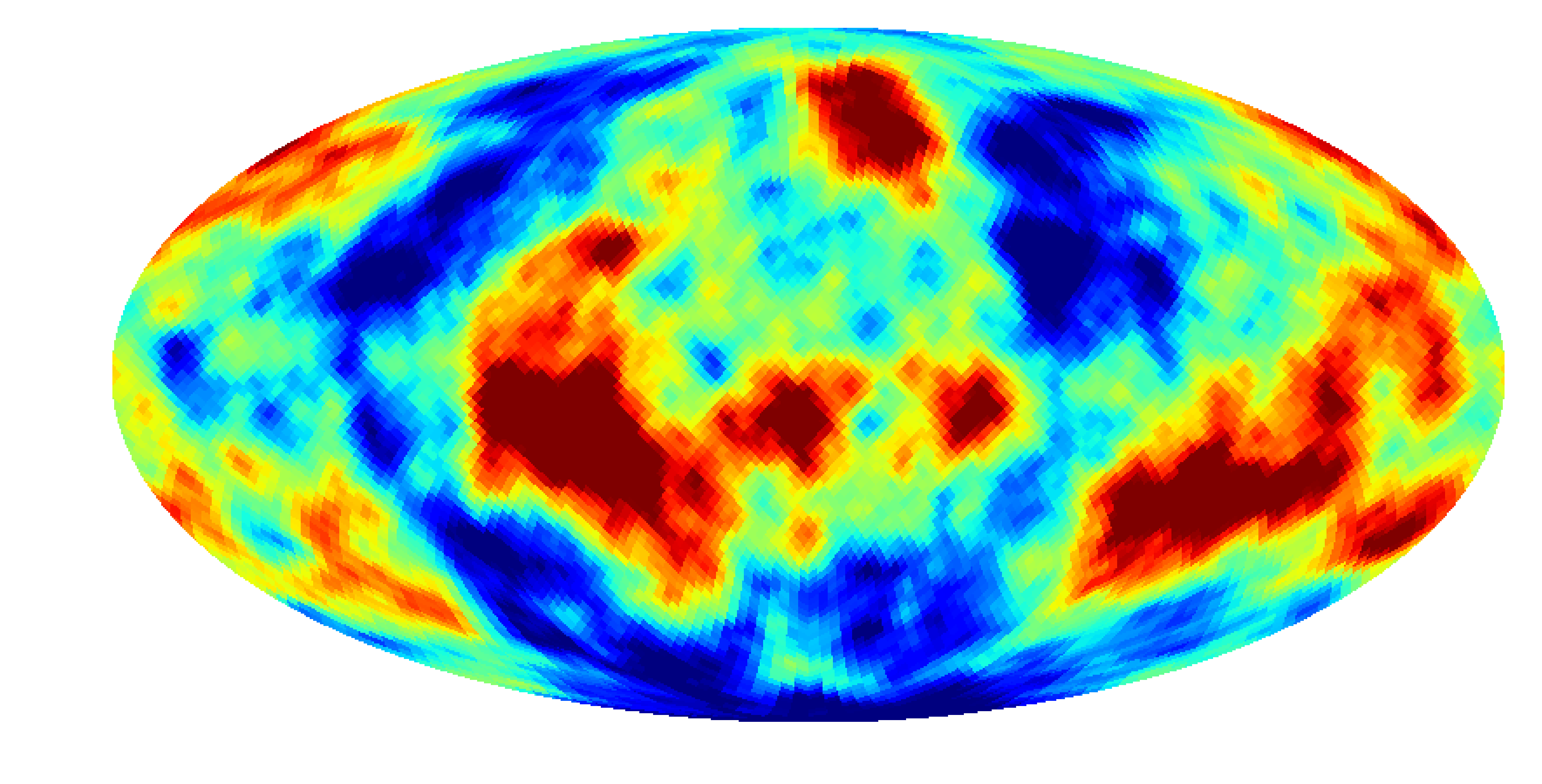}
\includegraphics[width=.4\textwidth]{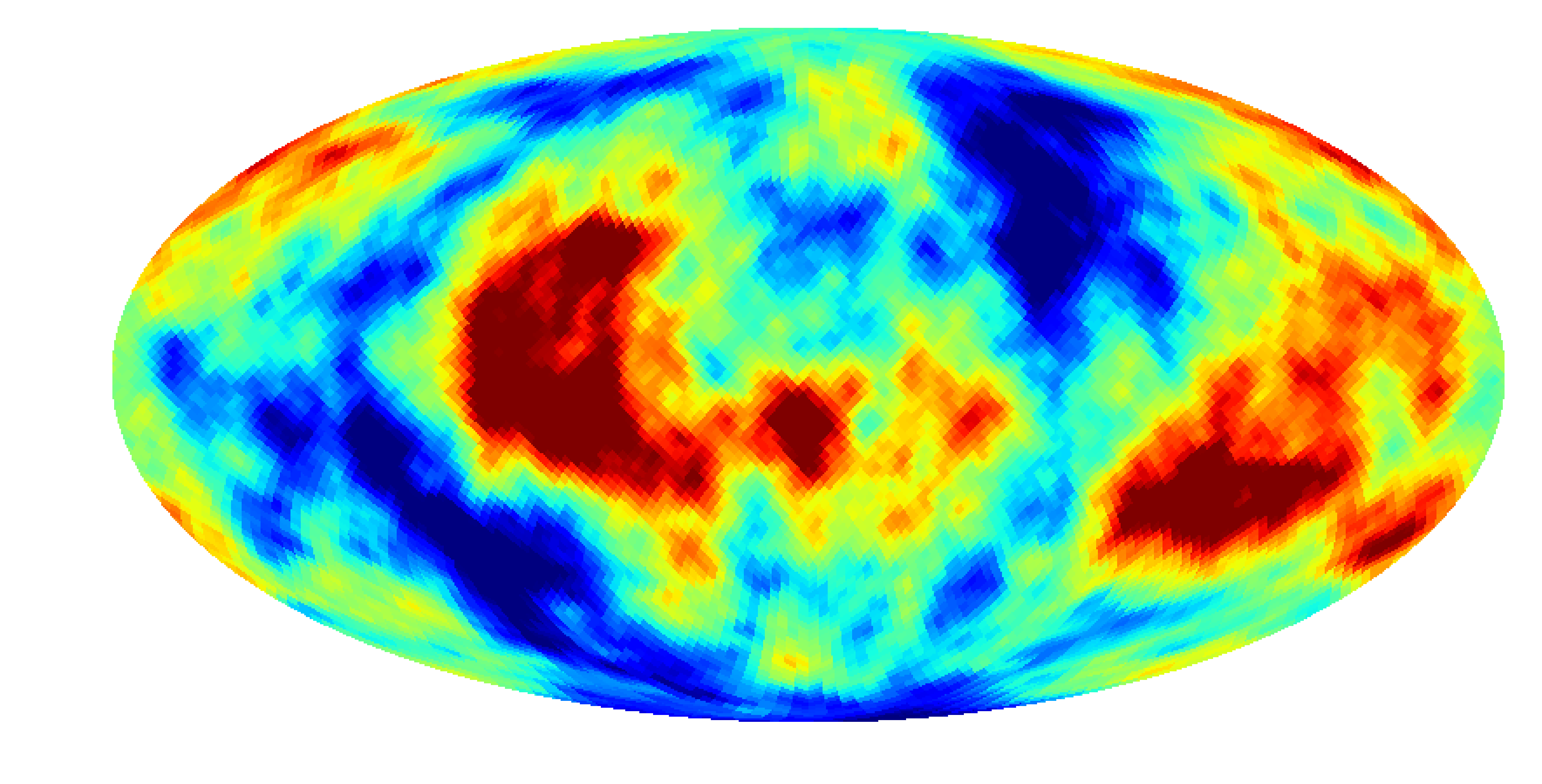}
\includegraphics[width=.4\textwidth]{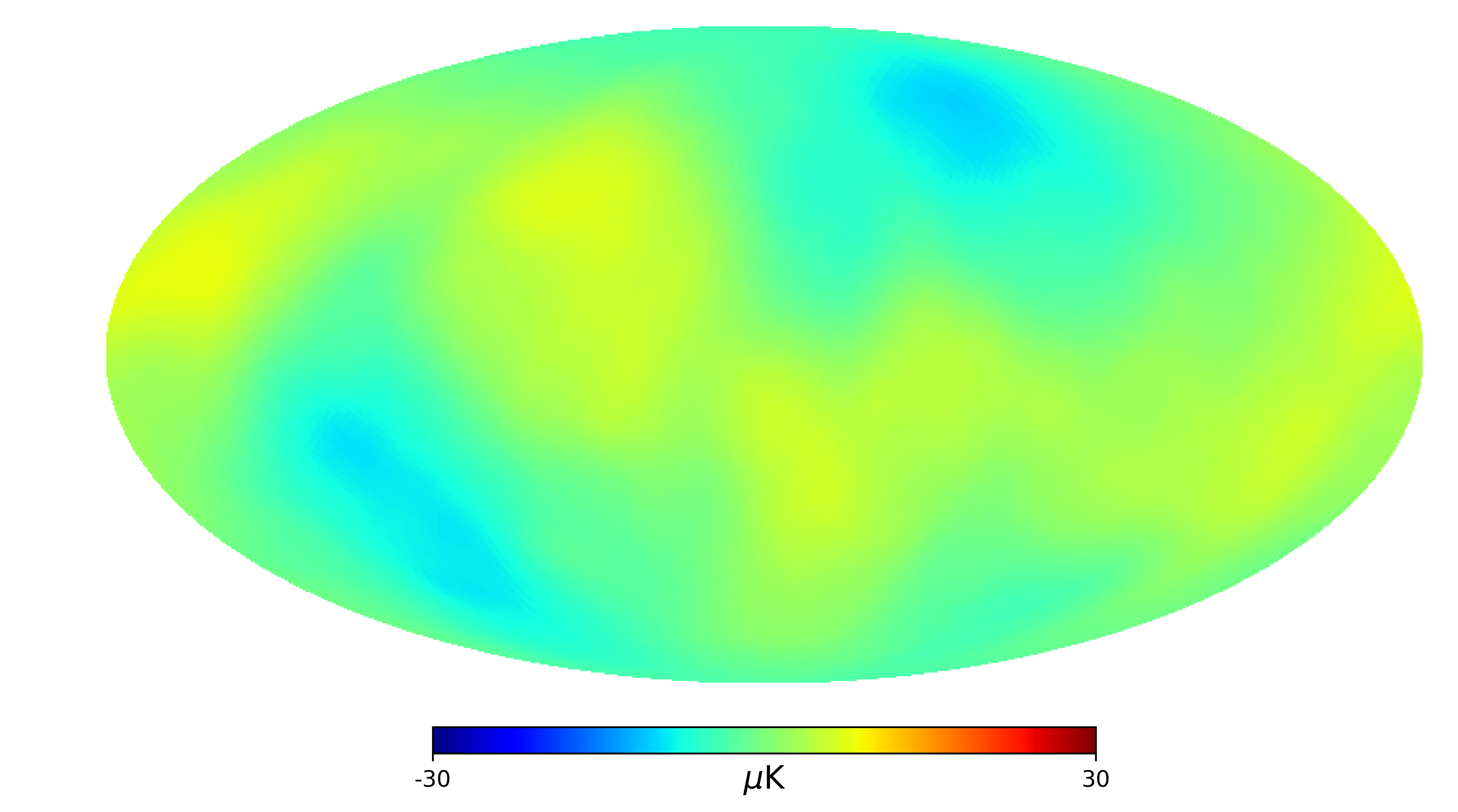}
\caption{Example of the reconstruction procedure applied to simulation data, see \eqref{eqn:estimator}. {\em Top:} input ISW map; {\em centre:}  
recovered ISW estimate using CMB temperature and SKA1 continuum  
($S_{\rm cut} = 22\,\mu$Jy); {\em bottom:} recovered ISW estimate using 
CMB temperature alone. (All maps have resolution $N_{\rm side} = 32$.)}
\label{fig:maps}
\end{figure}

Following \cite{Manzotti:2014kta,Bonavera:2016hbm}, we show for illustrative purposes an 
example of the reconstructed ISW map in figure~\ref{fig:maps}.
The reconstruction using  simulated data  of a SKA1-MID Band 1 radio continuum survey with 
$S_{\rm cut} = 22$ $\mu$Jy with 5 $z$-bins (corresponding to a SNR $\sim 5.0$) (central panel), 
shares a number of the visual features in common with the input ISW map (top panel). For comparison, 
we show in the bottom panel the ISW reconstruction obtained with the CMB temperature map alone 
(reducing \eqref{eqn:estimator} to a Wiener filter), where only a very large scale feature is captured.

\subsection{Reconstruction validation}
To quantify the accuracy of a given reconstruction, we use the correlation coefficient between 
the true ISW map $T^{\rm ISW}$ and the reconstructed ISW map $\hat{T}^{\rm ISW}$:
\be
\label{eqn:rho}
\rho = \frac{\big\langle T^{\rm ISW}\hat{T}^{\rm ISW} \big\rangle_{\rm pix}}{\sigma_{\rm ISW}\sigma_{\rm rec}} \,,
\ee
where $\sigma_{\rm ISW}$ ($\sigma_{\rm rec}$) is the variance of the true (reconstructed) ISW map.
We consider also a second statistical estimator, because the estimator \eqref{eqn:rho} is 
insensitive to changes in the overall amplitude of the reconstructed ISW map. 
The reconstructed map residual is defined as: 
\be
\label{eqn:s}
s = \frac{\big\langle \big(T^{\rm ISW}-\hat{T}^{\rm ISW}\big) \big\rangle_{\rm pix}^{1/2}}{\sigma_{\rm ISW}} \,.
\ee

\begin{table*}
\centering
\caption{Mean reconstruction quality coefficients $\langle \rho \rangle$ and $\langle s \rangle$ 
of ISW map reconstructions for various combinations of input maps. First column gives the case 
where GR corrections are included in the estimator. 
Second column shows results using the Newtonian approximation. Numbers in brackets indicate that 
cross-bin correlations are {\em neglected}.}
\begin{tabular}{|c|c|c|c|c|}
\hline
\rule[-1.2mm]{0mm}{.45cm}
 & \multicolumn{2}{|c|}{$\langle \rho \rangle$}  & \multicolumn{2}{|c|}{$\langle s \rangle$} \\
\hline
\rule[-1.2mm]{0mm}{.45cm}
 & $R_\ell(\Delta^{\rm N+GR}_\ell)$ & $R_\ell(\Delta^{\rm N}_\ell)$ & $R_\ell(\Delta^{\rm N+GR}_\ell)$ & $R_\ell(\Delta^{\rm N}_\ell)$\\
\hline\hline
\rule[-1.2mm]{0mm}{.45cm}
MeerKLASS L-Band (1 bin) & 0.38 & 0.38  & 0.93 & 0.93 \\
\rule[-1.2mm]{0mm}{.45cm}
MeerKLASS L-Band (5 bin) & 0.54 (0.43) & 0.54 (0.43) & 0.84 (0.91) & 0.84 (0.90) \\
\hline
\rule[-1.2mm]{0mm}{.45cm}
MeerKLASS UHF-Band (1 bin) & 0.56 & 0.56  & 0.83 & 0.83 \\
\rule[-1.2mm]{0mm}{.45cm}
MeerKLASS UHF-Band (11 bin) & 0.65 (0.57) & 0.65 (0.41) & 0.76 (0.87) & 0.77 (0.91) \\
\hline
\rule[-1.2mm]{0mm}{.45cm}
SKA1 - IM (1 bin) & 0.55 & 0.55  & 0.85 & 0.85 \\
\rule[-1.2mm]{0mm}{.45cm}
SKA1 - IM (27 bin) & 0.75 (0.66) & 0.75 (0.43) & 0.63 (0.85) & 0.65 (0.90) \\
\hline\hline
\rule[-1.2mm]{0mm}{.45cm}
EMU (1 bin) & 0.63 & 0.62 & 0.78 & 0.78 \\
\hline
\rule[-1.2mm]{0mm}{.45cm}
SKA1 - continuum 22 $\mu$Jy (1 bin) & 0.67 & 0.67  & 0.75 & 0.75 \\
\rule[-1.2mm]{0mm}{.45cm}
SKA1 - continuum 22 $\mu$Jy (5 bin) & 0.87 (0.78)  & 0.87 (0.76) & 0.53 (0.65) & 0.49 (0.65) \\
\hline
\rule[-1.2mm]{0mm}{.45cm}
SKA1 - continuum 10 $\mu$Jy (1 bin) & 0.93 & 0.93  & 0.39 & 0.37 \\
\rule[-1.2mm]{0mm}{.45cm}
SKA1 - continuum 10 $\mu$Jy (5 bin) & 0.93 (0.85) & 0.93 (0.82) & 0.41 (0.55) & 0.36 (0.57) \\
\hline\hline
\rule[-1.2mm]{0mm}{.45cm}
SKA2 - continuum 1 $\mu$Jy (1 bin) & 0.82 & 0.82  & 0.58 & 0.60 \\
\rule[-1.2mm]{0mm}{.45cm}
SKA2 - continuum 1 $\mu$Jy (5 bin) & 0.90 (0.77) & 0.88 (0.84) & 0.43 (0.64) & 0.54 (0.61) \\
\hline
\rule[-1.2mm]{0mm}{.45cm}
SKA2 - HI gal (1 bin) & 0.60 & 0.60  & 0.44 & 0.49 \\
\rule[-1.2mm]{0mm}{.45cm}
SKA2 - HI gal (19 bin) & 0.90 (0.67) & 0.88 (0.53) & 0.44 (0.75) & 0.49 (0.85) \\
\hline
\end{tabular}
\label{tab:rho}
\end{table*}

We calculate $\rho$ and $s$, averaged over 10,000 simulations, using the following general pipeline:
\begin{itemize}
\item Fiducial cosmological model is flat, with best-fit parameters: 
$\omega_\textrm{b}\equiv\Omega_\textrm{b}h^2=0.02218$, 
$\omega_\textrm{c}\equiv\Omega_\textrm{c}h^2=0.1205$, $h_0=0.6693$, $\tau=0.0596$, 
$n_\textrm{s}=0.9619$, and $\log\left(10^{10}\ A_\textrm{s}\right)=3.056$,
from the 2015 analysis of $Planck$ data \citep{Aghanim:2016yuo}.
LSS survey specifications are given in section~\ref{sec:four}.
\item Compute the observed auto- and cross-correlation angular power spectra \eqref{eqn:APS}, 
including the GR corrections with a modified version of 
CAMB\_sources\footnote{\href{https://github.com/cmbant/CAMB/tree/CAMB\_sources}{https://github.com/cmbant/CAMB/tree/CAMB\_sources}} \citep{Challinor:2011bk}.
\item Generate correlated Gaussian realisations of the CMB and LSS maps using HEALPix\footnote{\href{https://github.com/healpy/healpy}{https://github.com/healpy/healpy}} 
\citep{Gorski:2004by}:
\be
a_{\ell m}^{i} = \sum_{j=1}^{i} \xi_{j} T_{ij} \,,
\ee
where $\xi$ is a complex random number with unit variance $\langle \xi \xi^* \rangle = 1$ 
and zero mean $\langle \xi \rangle = 0$,  satisfying $\langle \xi_i \xi^*_j \rangle = \delta_{ij}$. 
The amplitudes $T_{ij}$ are generated with the following recursive expression 
\citep{Giannantonio:2008zi} to guarantee that any two maps will be correlated:
\begin{align}
&T_{ij} = \left({C^{ji}-\sum_{k=1}^{j-1}T_{ik}^2}\right)^{1/2} \qquad\qquad~~ \text{ if  }~ i=j\,, \\
&T_{ij} =\big( T_{jj}\big)^{-1} \left({C^{ji}-\sum_{k=1}^{j-1}T_{ik}T_{jk}}\right) \quad \text{ if  }~ i>j \,,
\end{align}
where the index $i$ runs over the number of maps (CMB and LSS).
\item The noise for the LSS maps, i.e. shot-noise for the galaxy surveys and instrumental 
noise for IM, is also generated in the form of $a_{\ell m}^{\cal N}$ as a Gaussian uncorrelated 
map with respect to the other fields.
\item Construct the covariance \eqref{eqn:cov} with a set of auto- and cross-correlation 
angular power spectra $C_{\ell}^{\rm XY}$. As we did for the SNR estimation, we test both the 
impact of including the GR corrections (always included in the input maps/ simulations) and 
neglecting cross-bin correlations in the covariance \eqref{eqn:cov}.
\item Compare the reconstructed ISW signal to the true ISW map and evaluate the accuracy of 
the reconstruction by using \eqref{eqn:rho} and \eqref{eqn:s}.
\end{itemize}

Table~\ref{tab:rho} shows the quality of the ISW reconstruction for different datasets, 
where values  $\langle \rho \rangle \to 1$ and $\langle s \rangle \to 0$ indicate more 
accurate reconstruction.

We consider full-sky simulations without taking into account the mask for each single map.
\cite{Bonavera:2016hbm} show that the reconstruction quality is degraded by incomplete sky 
coverage input datasets, even when considering spectra corrected using MASTER \citep{Hivon:2001jp} 
in order to include the mode coupling in the presence of a mask.

The values of $\langle\rho\rangle$, averaged over 10,000 simulations, follow our finding for 
the SNR. We find always a higher value of $\langle\rho\rangle$ when tomography is performed, 
even in those cases, like MeerKLASS and SKA1 continuum surveys, where there is a tiny improvement 
in terms of SNR.
Neglecting the GR terms in the filter for the reconstruction, by using a wrong covariance matrix 
\eqref{eqn:cov}, the recovered map is not properly scaled by the filter and the quality of the 
reconstruction is worse. However including the cross-correlation between redshift bins 
compensates for this degradation and we find the same values for $\langle \rho \rangle$, even 
when GR terms are not included in the theoretical angular power spectra used for the covariance.
The inclusion of the cross-correlation leads to a better reconstruction quality, 
around $\sim 10 - 20\%$. 

Similar findings follow for the average reconstructed map residuals $\langle s \rangle$.
The correlation between the galaxy number count contrast and the ISW component of the CMB 
is affected  both in shape and amplitude at  large angular scales, by the lensing convergence contribution to the number counts (see figures~\ref{fig:Tg_cont4} and \ref{fig:Tg_gal4}).
However, this effect becomes important at high redshift $z>1.5$, which lowers the contribution 
to the total ISW detection -- and for this reason we do not see a significant shift 
in $\langle s \rangle$ when the wrong covariance matrix is used.

\section{Conclusions}
\label{sec:conclusion}
In this work, we studied the feasibility of detecting the late-time ISW imprinted on the CMB 
temperature anisotropies by cross-correlating CMB maps with future radio maps from the SKA. 
Then we investigated the reconstruction of the ISW signal combining CMB with SKA surveys. 
We considered two of the main three cosmological probes provided by SKA in Phase 1 
\citep{Bacon:2018dui}, namely the neutral hydrogen (HI) intensity mapping survey and the radio 
continuum galaxy survey, together with their two precursor surveys, MeerKLASS \citep{Santos:2017qgq} 
and EMU \citep{Norris:2011ai}.
We also considered the more futuristic SKA2 for radio continuum and the HI galaxy survey (the 
`billion galaxy' spectroscopic survey).

We began by quantifying the theoretical signal-to-noise ratio (SNR) for ISW detection through 
angular cross-power spectra of CMB temperature and number count/ intensity contrast.
One of the key factors to maximize the synergy between CMB and LSS maps is to have the largest 
overlapping sky area, also because SNR $\propto \sqrt{f_{\rm sky}}$. Future radio surveys promise 
to deliver maps of the dark matter distribution using HI with sky area $\sim 20,000-30,000\,$deg$^2$ 
so that the surveys are signal-dominated on the large scales relevant for the ISW. On the other 
hand, future optical/ infrared surveys such as DESI \citep{Levi:2013gra,Aghamousa:2016sne,Aghamousa:2016zmz}, 
Euclid \citep{Laureijs:2011gra,Amendola:2016saw} 
and LSST \citep{Abell:2009aa,Mandelbaum:2018ouv} have sky area $\sim 15,000\,$deg$^2$.
The deeper redshift coverage of the radio surveys also increases the SNR, but not 
significantly, as shown in figure~\ref{fig:SNR}. This is due to the fact that matter-temperature 
correlations decrease rapidly with redshift, since dark energy is subdominant at high redshift.

For the HI IM surveys, we find for MeerKLASS $1 < \text{SNR} <2$ ($\sim 2.2$ combining the two 
Bands) and SNR $\sim 3.7-4.7$ (higher with tomography) for SKA1 in Band 1: see table~\ref{tab:SNim}. 
For the radio continuum galaxy surveys, we find for EMU that SNR $\sim 5$, while SNR $\sim 4.0-5.0$ 
(higher with tomography) for SKA1 in Band 1 and $5.6-6.2$ (higher with tomography) for SKA2: see 
table~\ref{tab:SNcont}. For the HI galaxy survey from SKA2 we find a SNR $\sim 3.7-6.0$ (higher with 
tomography): see table~\ref{tab:SNgal}.

We tested the effect on the SNR of a tomographic approach, splitting the information of 
the surveys in different redshift bins. The results show that tomography does improve the SNR.

Then we considered the reconstruction of the ISW signal using a likelihood based minimum-variance 
estimator from CMB and LSS maps. 
We considered the ISW reconstruction from single radio surveys but splitting the surveys in 
different redshift bins, according to their baseline specifications and consistently with our SNR analysis.

We quantifed the accuracy of the reconstruction with two estimators, namely 
the correlation coefficient between the input and the reconstructed ISW map $\rho$,
and the reconstructed map residual $s$. 
We calculated $\rho$ and $s$, averaging over 10,000 simulations, for all the radio surveys 
considered, with and without tomography.
Our results for $\langle \rho \rangle$ and $\langle s \rangle$ are consistent with our finding 
for the SNR. Moreover, we find that tomography always leads to a higher quality of the ISW 
reconstruction for these two estimators.

We also studied the impact of observational effects on the radio surveys from lensing magnification, 
Doppler and other relativistic corrections, in altering the cross-correlation signal from CMB 
temperature and radio surveys. 
We found that the lensing effects do alter the angular cross-power spectra of CMB temperature 
and number count/ intensity contrast, as seen in figures~\ref{fig:Tg_im4}--\ref{fig:Tg_gal4}, 
while the other relativistic effects can be neglected.  
Lensing magnification can change the expected SNR up to 10-20\% and degrade the ISW reconstruction 
when it is not modelled in the theoretical covariance used in \eqref{eqn:SNR} for the SNR and 
used in \eqref{eqn:estimator} for the reconstruction. 


One of our main results is to show the {\em importance of including cross-bin correlations of 
the LSS survey} in computing the SNR. These correlations enter the covariance \eqref{covm}, 
and neglecting this contribution leads to significant bias in the SNR, as shown by the dashed 
curves in figure~\ref{fig:SNR} and by the bracketed numbers in tables~\ref{tab:SNim}--\ref{tab:SNgal}. When the cross-bin galaxy correlations are neglected, there is also a significant disagreement 
between the SNR in the Newtonian approximation and in the full GR model (i.e. including lensing). 
This is evident in figure~\ref{fig:SNR} and in tables~\ref{tab:SNcont} and \ref{tab:SNgal}.

We showed that for the galaxy surveys, when cross-bin correlations are {\em included,} the 
SNR in the Newtonian approximation (i.e. no lensing) differs negligibly from the SNR in 
the correct model (full GR, including lensing). This implies that lensing contributions 
within bins and across different bins combine to effectively cancel [see \eqref{n=gr}].

We conclude that SKA in Phase 1 promises a $\sim 5\sigma$ detection of the ISW signal 
with 21cm intensity mapping and radio continuum surveys, with a similar forecast for the 
precursor EMU survey, while a larger significance at $\sim 6\sigma$ will be possible 
with SKA2 using the 21cm galaxy redshift and radio continuum surveys.
Moreover, we find that lensing and other relativistic observation effects on the number 
counts/ intensity angular power spectra have a small  impact on the ISW detection and reconstruction.
However, their impact on cosmological parameter estimation can be significant \citep{Camera:2014bwa,Cardona:2016qxn,Lorenz:2017iez}.

There are a number of ways that one could improve the robustness and accuracy of our forecasts.

We considered the cross-correlation between the CMB and single radio surveys.
Since the estimator \eqref{eqn:estimator} is able to combine any numbers of maps as input, 
it is possible to test the combined effect of all the radio surveys that will be provided from SKA 
for the ISW detection and reconstruction.
However, this multi-tracer application to the ISW has been shown in \cite{Ballardini:2017xnt} 
to be effective for surveys which cover different redshifts, or for tracers with a very different 
redshift distribution. 

The inclusion of maps of the lensing potential has been shown 
\citep{Cooray:2001ab,Ferraro:2014msa,Manzotti:2014kta,Ade:2015dva,Bonavera:2016hbm} 
to have the potential to improve the ISW reconstruction.  
This will be possible in light of future CMB experiments beyond $Planck$, able to 
provide better lensing maps in particular at the largest scales, i.e. $\ell < 10$, 
where the correlation between ISW and lensing is largest \citep{Manzotti:2014kta}.

Finally, we note that systematics will impact the ISW detection significances forecasted here. 
Estimation of these systematics is a major undertaking, since no cosmological radio survey has 
yet been implemented, and the 3 different radio surveys studied here will be affected by 
different systematics, requiring dedicated treatments.  
Further investigation can build on previous work on optical/IR surveys, for example 
\cite{Afshordi:2004kz,HernandezMonteagudo:2009fb,Bonavera:2016hbm,Muir:2016veb,Weaverdyck:2017ovf,Ballardini:2017xnt}.
In addition, systematics need to be taken into account in the reconstruction estimator 
to avoid biased results \citep{Muir:2016veb,Weaverdyck:2017ovf}.

\newpage
\subsection*{Acknowledgments}
MB thanks R.B. Barreiro, C.A.P. Bengaly, L. Bonavera, S. Camera, F. Finelli, 
J. Fonseca, A. Manzotti, D. Molinari, M. Spinelli and C. Umilt\`a for fruitful discussions.
MB and RM were supported by the South African Radio Astronomy Observatory, which is 
a facility of the National Research Foundation, an agency of the Department of Science and Technology (Grant No. 75415).
MB is also supported by a Claude Leon Foundation fellowship and by ASI n.I/023/12/0 
``Attivit\'a relative alla fase B2/C per la missione Euclid".
RM is also supported by the UK Science \& Technology Facilities Council (Grant ST/N000668/1).




\bibliographystyle{mnras}
\bibliography{Biblio} 







\bsp	
\label{lastpage}
\end{document}